\documentclass[aps,prd,reprint,nofootinbib,superscriptaddress,longbibliography]{revtex4-2}

\usepackage{amsmath,amssymb,bm}
\usepackage{graphicx}
\usepackage{booktabs}
\usepackage{multirow}
\usepackage{xcolor}
\usepackage{hyperref}
\usepackage{microtype}
\usepackage{placeins}

\hypersetup{colorlinks=true,citecolor=blue,urlcolor=blue,linkcolor=blue}
\newcommand{\kappam}{\kappa_{-}}
\newcommand{\kappap}{\kappa_{+}}
\newcommand{\kappac}{\kappa_{c}}
\newcommand{\qmax}{q_{\max}}
\newcommand{\betamin}{\beta_{\mathrm{min}}}
\newcommand{\qhat}{\widehat q}
\newcommand{\dd}{\mathrm{d}}

\begin{document}
\flushbottom

\renewcommand{\topfraction}{0.95}
\renewcommand{\dbltopfraction}{0.95}
\renewcommand{\textfraction}{0.05}
\renewcommand{\floatpagefraction}{0.90}
\renewcommand{\dblfloatpagefraction}{0.90}
\setcounter{topnumber}{4}
\setcounter{dbltopnumber}{3}
\setcounter{totalnumber}{6}
\setlength{\textfloatsep}{8pt plus 2pt minus 2pt}
\setlength{\dbltextfloatsep}{8pt plus 2pt minus 2pt}
\setlength{\floatsep}{6pt plus 2pt minus 2pt}

\title{Near-extremal asymptotics and strong cosmic censorship for black holes immersed in a Chaplygin-like dark fluid}

\author{Hao-Peng Yan}
\email[Corresponding author: ]{yanhaopeng@tyut.edu.cn}
\author{Zeng-Yi Zhang}
\author{Xiang-Qian Li}
\author{Xiao-Jun Yue}

\affiliation{College of Physics and Optoelectronic Engineering, Taiyuan University of Technology, Taiyuan 030024, China}

\begin{abstract}
We investigate strong cosmic censorship (SCC) for massless scalar perturbations of electrically neutral black holes immersed in a Chaplygin-like dark fluid (CDF).  The matter distribution produces an asymptotically de Sitter exterior while supporting an inner Cauchy horizon.  We derive a closed-form parametrization of the extremal and Nariai horizon boundaries.  This analytic control yields explicit near-extremal asymptotics for the horizon splitting and Cauchy-horizon surface gravity, together with an analytic expression for the leading near-extremal scalar quasinormal spectrum.  The fundamental near-extremal damping rate approaches the Cauchy-horizon surface gravity, whereas higher angular multipoles retain explicit dependence on the CDF extremal geometry.  Combining these analytic results with global quasinormal-mode calculations, we determine the spectral gap and assess the Christodoulou formulation of SCC throughout the three-horizon domain.  Potential SCC violation is confined to a narrow region near extremality.  As the effective cosmological scale increases, the lower boundary of this region is controlled successively by de Sitter, photon-sphere, and de Sitter modes, while its normalized width varies nonmonotonically.
\end{abstract}

\maketitle

\section{Introduction}
\label{sec:intro}

General relativity is expected to provide a deterministic evolution from suitable initial data. The presence of a Cauchy horizon challenges this expectation because the spacetime beyond it is not uniquely determined by data on the original Cauchy surface. Strong cosmic censorship (SCC) addresses this issue by requiring, in an appropriate regularity class, that the maximal globally hyperbolic development of generic initial data cannot be extended beyond the Cauchy horizon \cite{Penrose1969}. Perturbations falling into an inner horizon are strongly blueshifted, providing the physical basis for the instability and mass-inflation picture of charged black-hole interiors \cite{PoissonIsrael1990,Dafermos2003,Dafermos2005}. In asymptotically de Sitter spacetimes, however, exterior perturbations decay exponentially \cite{Brady1997}, and the competition between this decay and the inner-horizon blueshift can substantially weaken the singular behavior at the Cauchy horizon \cite{CostaFranzen2017,CostaEtAl2017}.

For a massless scalar field, the slowest exterior decay rate is characterized by the quasinormal-mode (QNM) spectral gap $\alpha$. It is convenient to introduce \cite{HintzVasy2017,Cardoso2018}
\begin{equation}
    \beta\equiv\frac{\alpha}{\kappam},
    \qquad
    \alpha=\inf_{\ell,n}\{-\operatorname{Im}\omega_{\ell n}\},
    \label{eq:beta_intro}
\end{equation}
where $\ell$ and $n$ denote the angular multipole and overtone numbers, respectively, and $\kappam$ is the Cauchy-horizon surface gravity. For smooth scalar initial data, $\beta>1/2$ corresponds to sufficient local Sobolev regularity at the Cauchy horizon to indicate a possible violation of the Christodoulou formulation of SCC \cite{HintzVasy2017,Cardoso2018}. Cardoso \emph{et al.} showed that this can occur for near-extremal Reissner--Nordstr\"om--de Sitter (RNdS) black holes and identified three QNM families relevant to the spectral gap: photon-sphere (PS), de Sitter (dS), and near-extremal (NE) modes \cite{Cardoso2018}. Subsequent studies have examined rotating de Sitter geometries and different perturbing fields, including charged scalar and fermionic fields \cite{DiasKdS2018,Mo2018,CardosoSubtle2018,DiasCharged2019,Ge2019,Destounis2019}. The QNM--SCC relation has also been investigated in higher-dimensional spacetimes and in backgrounds modified by nonminimal couplings, nonlinear electrodynamics, or alternative gravitational dynamics \cite{LiuHigherD2019,Guo2019,Gan2019,SangJiang2022,JiangTan2023,Shao2024,KonoplyaZhidenko2022}. The role of initial-data regularity, nonlinear matter dynamics, and quantum effects has been clarified further in Refs.~\cite{Dias2018,DafermosShlapentokh2018,Luna2019,LunaAddendum2021,KehleVanDeMoortel2024,Rossetti2025,HollandsWaldZahn2020,Chrysostomou2025}. Recent work continues to find that the outcome depends sensitively on the geometry and perturbing sector \cite{Courty2023,Davey2024,Lin2025,LiWangJing2026,LiuGuo2026}.

A related question arises for electrically neutral black holes whose Cauchy horizon is supported by matter rather than electric charge. Near-extremal SCC violation has been found for an uncharged accelerating black hole \cite{ZhangJiang2023}. A narrow near-extremal potential-violation region has also been found for a spherically symmetric de Sitter black hole surrounded by perfect-fluid dark matter (PFDM) \cite{Jiang2024}.

Black holes immersed in a Chaplygin-like dark fluid (CDF) provide a matter-supported setting in which to examine this competition.  The CDF density approaches a constant at large radius and therefore generates an asymptotically de Sitter geometry, while its inhomogeneous component can support an inner Cauchy horizon \cite{Li2023,Li2024}.  In contrast to models in which the cosmological scale is specified independently of the matter profile, varying the CDF background changes both the asymptotic and strong-field geometry.  The model therefore allows the SCC competition to be studied when the asymptotic decay scale and the strong-field geometry are controlled by the same matter sector.

The scalar QNM spectrum of the original CDF black hole, including the photon-sphere (PS), de Sitter (dS), and near-extremal (NE) families, was studied in Ref.~\cite{Becar2024}, while its horizon multiplicity and global causal structure were analyzed in Ref.~\cite{Fontana2026}.  Here we investigate the SCC problem for this geometry.  We first derive a unified closed-form parametrization of the Nariai and extremal (cold) degenerate-horizon branches.  The parametrization makes the degenerate-horizon geometry explicit and allows us to derive analytic near-extremal asymptotics for the horizon splitting and Cauchy-horizon surface gravity, together with the leading NE spectrum for general angular multipole number $\ell$.  The fundamental mode approaches $\beta_{\rm NE}=1$ in the cold limit, whereas higher multipoles retain explicit dependence on the extremal CDF geometry.  We then determine the global dS, PS, and NE spectral competition throughout the three-horizon domain.  The resulting SCC phase boundary exhibits a dS$\to$PS$\to$dS sequence in the mode controlling the SCC threshold and bounds a narrow potential-violation strip with a nonmonotonic normalized width.

The paper is organized as follows.  Section~\ref{sec:geometry} introduces the CDF black-hole geometry and its exact degenerate-horizon boundaries.  Section~\ref{sec:scc} reviews the massless-scalar perturbation problem and the SCC regularity criterion.  Section~\ref{sec:cold} derives the cold-limit geometry and near-extremal asymptotics.  Section~\ref{sec:results} presents the numerical method, QNM families, spectral-gap competition, SCC phase diagram, and a comparison with the PFDM case.  We summarize the main results in Sec.~\ref{sec:conclusion}; technical derivations and numerical convergence tests are collected in the appendices.

\section{Black holes immersed in a Chaplygin-like dark fluid}
\label{sec:geometry}

\begin{figure*}[!t]
    \centering
    \includegraphics[width=0.485\textwidth]{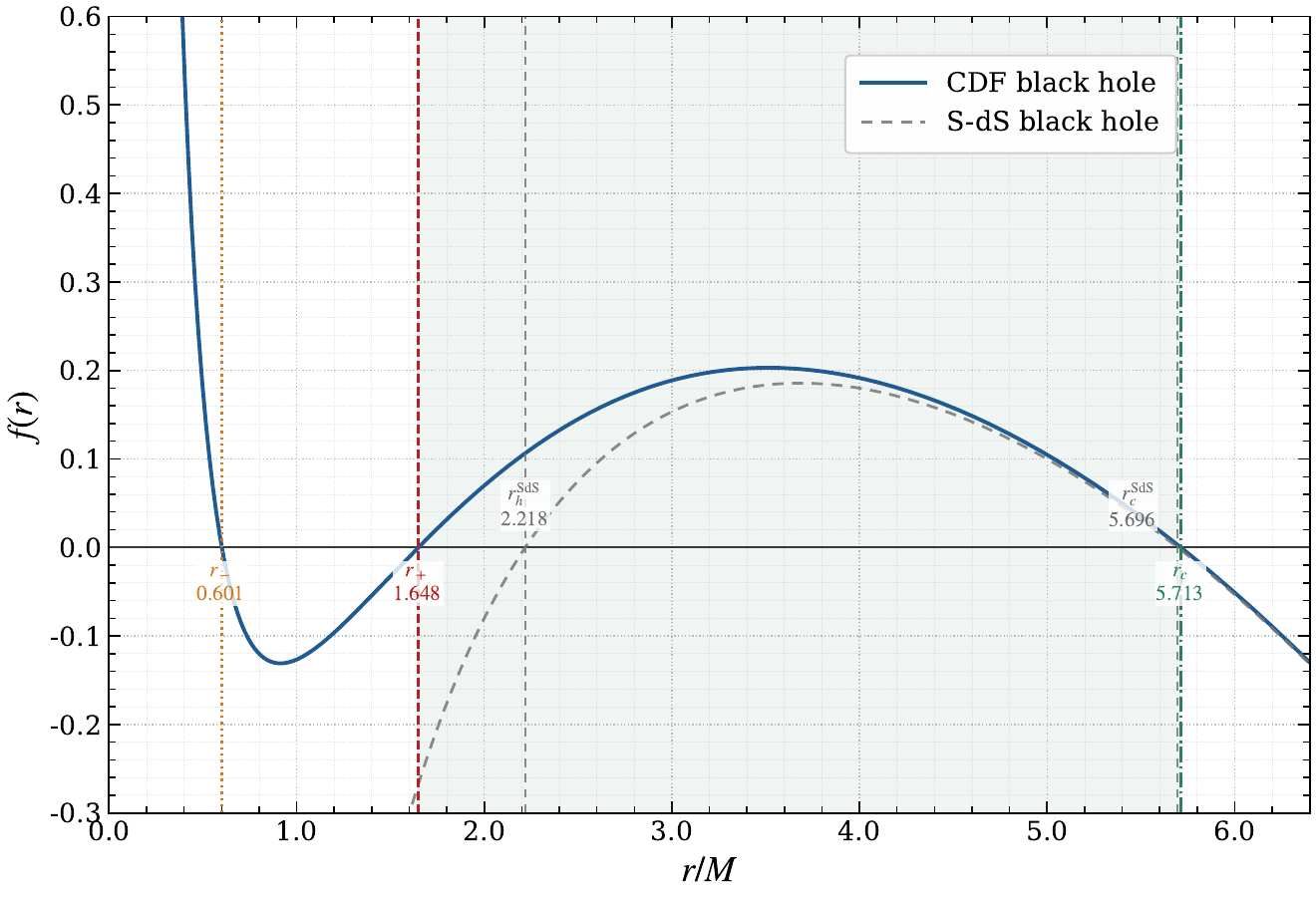}\hfill
    \includegraphics[width=0.485\textwidth]{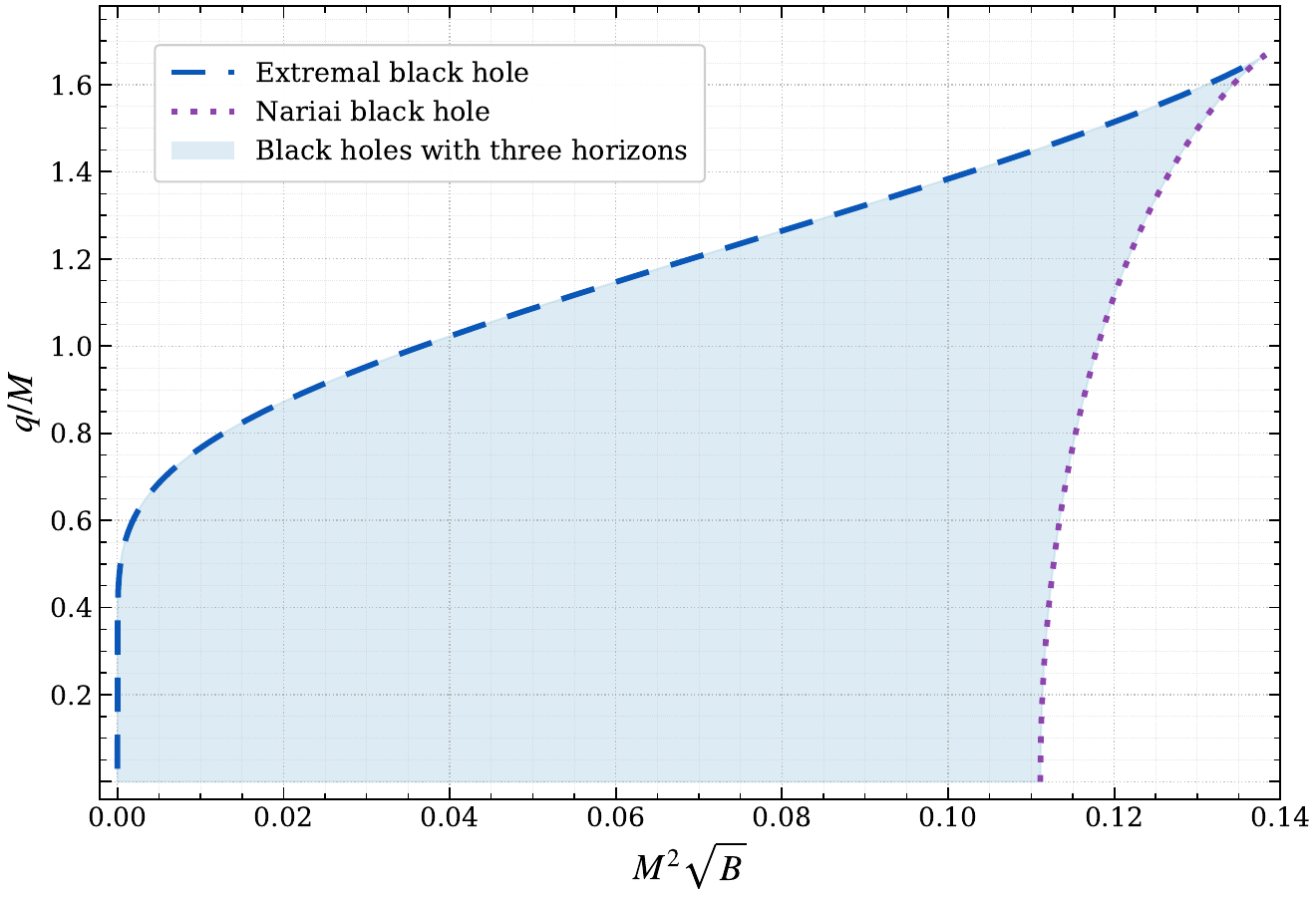}
    \caption{Geometry and three-horizon parameter domain of the CDF black hole. (a) Representative lapse function for $M=1$, $M^2\sqrt B=0.06$, and $q/\qmax=0.90$, with CDF horizons $r_-=0.601$, $r_+=1.648$, and $r_c=5.713$. The dashed curve is the Schwarzschild--de Sitter lapse function with the same mass and cosmological constant, $\Lambda M^2=M^2\sqrt B=0.06$, whose event and cosmological horizons are $r_h^{\rm SdS}=2.218$ and $r_c^{\rm SdS}=5.696$. (b) Parameter region for black holes with three horizons in the $(M^2\sqrt B,q/M)$ plane, bounded by the extremal curve $r_-=r_+$ and the Nariai curve $r_+=r_c$. The analytic boundaries are given by Eqs.~\eqref{eq:rdx}--\eqref{eq:Bdx} and meet at the triple-horizon point.}
    \label{fig:geometry}
\end{figure*}

We consider the static, spherically symmetric black-hole solution sourced by the Chaplygin-like dark fluid introduced in Refs.~\cite{Li2023,Li2024}.  The underlying Chaplygin-like relation is $p=-B/\rho$, with $B>0$.  In the static black-hole realization the effective source is anisotropic, with radial pressure $p_r=-\rho$, while the angularly averaged pressure satisfies the relation above.  Related black-hole environments based on modified Chaplygin-like fluids, including string-cloud extensions, have been investigated in Refs.~\cite{LiMCDF2025,Yan2026}.  The corresponding line element is
\begin{equation}
    \dd s^2=-f(r)\dd t^2+\frac{\dd r^2}{f(r)}+r^2\dd\Omega_2^2,
    \label{eq:metric}
\end{equation}
and the energy density and lapse function are \cite{Li2024}
\begin{equation}
    \rho(r)=\sqrt{B+\frac{q^2}{r^6}},
    \label{eq:rho}
\end{equation}
\begin{equation}
\begin{split}
    f(r)=1-\frac{2M}{r}
    &-\frac{r^2}{3}\sqrt{B+\frac{q^2}{r^6}} \\
    &+\frac{q}{3r}\operatorname{arcsinh}
       \left(\frac{q}{\sqrt B\,r^3}\right),
\end{split}
    \label{eq:f}
\end{equation}
The two CDF parameters characterize complementary parts of the same density profile: $B$ fixes the asymptotic density floor, while $q>0$ controls the inhomogeneous radial component.  In the region where the latter dominates, $\rho\sim q/r^3$, while at large radius
\begin{equation}
    \rho(r)\to\sqrt B,
    \qquad
    f(r)\to1-\frac{\sqrt B}{3}r^2,
    \label{eq:asymptotic}
\end{equation}
so the asymptotic geometry is de Sitter with effective cosmological constant $\Lambda_{\rm eff}=\sqrt B$.  Since $M$ sets the length scale, we present the model parameters through the dimensionless combinations $M^2\sqrt B$ and $q/M$ and set $M=1$ in numerical calculations.  Figure~\ref{fig:geometry}(a) compares a representative CDF black hole with a Schwarzschild--de Sitter (SdS) black hole of the same mass and cosmological constant, $\Lambda M^2=M^2\sqrt B=0.06$.  The cosmological horizons are nearly coincident, while the CDF shifts the event horizon inward and introduces an additional Cauchy horizon.  Thus the CDF reproduces the same asymptotic de Sitter curvature while modifying the strong-field horizon structure relevant to SCC.

In the three-horizon domain the positive roots obey
\begin{equation}
    r_-<r_+<r_c,
    \label{eq:horizons}
\end{equation}
with $r_-$, $r_+$, and $r_c$ the Cauchy, event, and cosmological horizons, respectively, and
\begin{equation}
    \kappa_i=\frac12\left|f'(r_i)\right|,
    \qquad i\in\{-,+,c\}.
    \label{eq:kappa}
\end{equation}
Ref.~\cite{Fontana2026} showed that the CDF geometry admits at most three positive horizons and determined its global causal structure and critical bounds. The three-horizon domain is bounded by two degenerate configurations: the Nariai limit $r_+=r_c$ and the extremal (cold) limit $r_-=r_+$. Denoting the degenerate-horizon radius and the corresponding boundary parameters by $r_d$, $q_d$, and $B_d$, respectively, both boundaries can be parametrized exactly as derived in Appendix~\ref{app:horizon},
\begin{align}
    r_d(x)&=\frac{6M}{2+x\tanh x},\label{eq:rdx}\\
    q_d(x)&=r_d(x)\tanh x,\label{eq:qdx}\\
    B_d(x)&=\frac{\operatorname{sech}^2x}{r_d(x)^4},\label{eq:Bdx}
\end{align}
with $x\in[0,\infty)$. The two branches meet at the triple-horizon (ultracold) point
\begin{equation}
    x_\star=\operatorname{arctanh}\sqrt{\frac23}.
    \label{eq:branchsummary}
\end{equation}
For $0\le x<x_\star$ the parametrization gives the Nariai boundary, while $x>x_\star$ gives the extremal boundary.  For each fixed $B$ in the three-horizon domain, this extremal branch provides the upper boundary in $q$; we denote the corresponding boundary value by $\qmax(B)$.  Equivalently, $q=\qmax(B)$ is the configuration for which $r_-=r_+$.  Because the boundary data are explicit functions of $x$, the same parametrization also supplies the cold geometric quantities entering the near-extremal analysis in Sec.~\ref{sec:cold}.

\section{Massless scalar perturbations and the SCC criterion}
\label{sec:scc}

We consider a minimally coupled massless scalar test field and neglect its backreaction on the CDF geometry. The field satisfies
\begin{equation}
    \nabla_a\nabla^a\Phi=0.
    \label{eq:KG}
\end{equation}
With the separation
\begin{equation}
    \Phi(t,r,\theta,\phi)=
    \sum_{\ell,m}e^{-i\omega t}Y_{\ell m}(\theta,\phi)\frac{\psi(r)}{r},
    \label{eq:separation}
\end{equation}
the radial equation becomes
\begin{equation}
    \frac{\dd^2\psi}{\dd r_*^2}
    +\left[\omega^2-V_\ell(r)\right]\psi=0,
    \qquad \frac{\dd r_*}{\dd r}=\frac1{f(r)},
    \label{eq:radial}
\end{equation}
where
\begin{equation}
    V_\ell(r)=f(r)\left[
    \frac{\ell(\ell+1)}{r^2}+\frac{f'(r)}{r}
    \right].
    \label{eq:potential}
\end{equation}
Between the event and cosmological horizons the QNM boundary conditions are purely ingoing at $r_+$ and purely outgoing at $r_c$,
\begin{equation}
\begin{aligned}
    \psi&\sim e^{-i\omega r_*}, && r\to r_+,\\
    \psi&\sim e^{+i\omega r_*}, && r\to r_c.
\end{aligned}
    \label{eq:QNMbc}
\end{equation}
These conditions select a discrete set $\omega_{\ell n}$.  For $\ell=0$, the exactly static solution $\omega=0$ corresponds to a constant scalar field and does not describe dynamical decay; it is therefore omitted when identifying the spectral gap.

The relation between the exterior QNM decay and the Cauchy-horizon regularity follows from the local behavior near $r=r_-$. Since
\begin{equation}
    f(r)\simeq -2\kappam(r-r_-),
    \qquad
    r_*\simeq-\frac{1}{2\kappam}\ln|r-r_-|,
    \label{eq:innerlocal}
\end{equation}
the potentially non-smooth branch of a separated mode has the characteristic radial factor
\begin{equation}
    \Phi_{\rm sing}\propto |r-r_-|^{i\omega/\kappam}.
    \label{eq:singbranch}
\end{equation}
Its differentiability is controlled by $-\operatorname{Im}\omega/\kappam$. For generic smooth initial data, the slowest exterior decay is set by the spectral gap $\alpha$, leading to the regularity exponent \cite{HintzVasy2017,Cardoso2018}
\begin{equation}
    \beta=\frac{\alpha}{\kappam},
    \qquad
    \alpha=\inf_{\ell,n}\{-\operatorname{Im}\omega_{\ell n}\}.
    \label{eq:beta}
\end{equation}
If $\beta>1/2$, the scalar field is locally $H^1$ at the Cauchy horizon, indicating a potential violation of the Christodoulou formulation of SCC at the linear scalar level. Conversely, $\beta\le1/2$ is compatible with the expected loss of $H^1$ regularity. We therefore use
\begin{equation}
    \beta>\frac12
    \label{eq:SCCcriterion}
\end{equation}
as the diagnostic for potential SCC violation, following Refs.~\cite{Cardoso2018,Jiang2024}. This criterion applies to the linear massless-scalar sector considered here; the coupled metric--CDF problem and nonlinear backreaction are beyond the scope of the present analysis.

The cold limit is particularly important for this diagnostic.  As $r_-\to r_+$, the Cauchy-horizon surface gravity $\kappam$ tends to zero, so the SCC ratio becomes sensitive to how the damping rates of the relevant QNM families scale with $\kappam$.  The exact cold-boundary parametrization obtained in Sec.~\ref{sec:geometry} makes this near-extremal scaling analytically accessible for the CDF geometry.

\section{Near-extremal asymptotics and near-horizon modes}
\label{sec:cold}

The exact cold-boundary parametrization allows both the extremal background and its leading departure from extremality to be treated analytically.  On the cold branch,
\begin{equation*}
 r_-=r_+\equiv r_e,
 \qquad
 q=q_e\equiv \qmax(B).
\end{equation*}
Approaching this branch at fixed $B$, we write
\begin{equation}
 q=q_e(1-\epsilon),\qquad 0<\epsilon\ll1.
 \label{eq:epsilon}
\end{equation}
Let $x_e$ denote the cold-branch parameter determined by $B_d(x_e)=B$, and define
\begin{equation}
 u_e\equiv\frac{q_e}{r_e}=\tanh x_e,\quad r_e=\frac{6M}{2+x_eu_e},\quad
 D_e=3u_e^2-2>0.
 \label{eq:colddefs}
\end{equation}
At fixed $B$, the derivatives of the lapse at the extremal point are
\begin{equation}
 \left.\partial_r^2 f\right|_e=\frac{D_e}{r_e^2},\qquad
 \left.\partial_q f\right|_e=\frac{x_e}{3r_e},
 \label{eq:coldderivatives}
\end{equation}
where $|_e$ denotes evaluation at $(r,q)=(r_e,q_e)$.  Thus $D_e=r_e^2(\partial_r^2f)_e$ measures the dimensionless curvature of the lapse at the cold double root.  Expanding the horizon equation $f(r,q)=0$ about $(r_e,q_e)$ at fixed $B$, with $f|_e=(\partial_r f)|_e=0$, gives
\begin{align}
 r_\pm&=r_e\pm C_r(B)\sqrt\epsilon+O(\epsilon),
 \label{eq:rpmexpansion}\\
 C_r(B)&=r_e\sqrt{\frac{2u_ex_e}{3D_e}},
 \label{eq:Cr}\\
 \kappa_\pm&=K(B)\sqrt\epsilon+O(\epsilon),
 \label{eq:kappaexpansion}\\
 K(B)&=\frac{1}{2r_e}\sqrt{\frac{2u_ex_eD_e}{3}}.
 \label{eq:Kcoefficient}
\end{align}
Thus the same cold-boundary data that determine $\qmax(B)$ fix both the horizon-splitting coefficient $C_r(B)$ and the surface-gravity coefficient $K(B)$; the latter sets the leading Cauchy-horizon blueshift scale.  As $D_e\to0$ at the ultracold endpoint, the square-root expansion becomes nonuniform and the triple-root scaling requires a separate treatment.

\begin{figure*}[!t]
 \centering
 \includegraphics[width=0.99\textwidth]{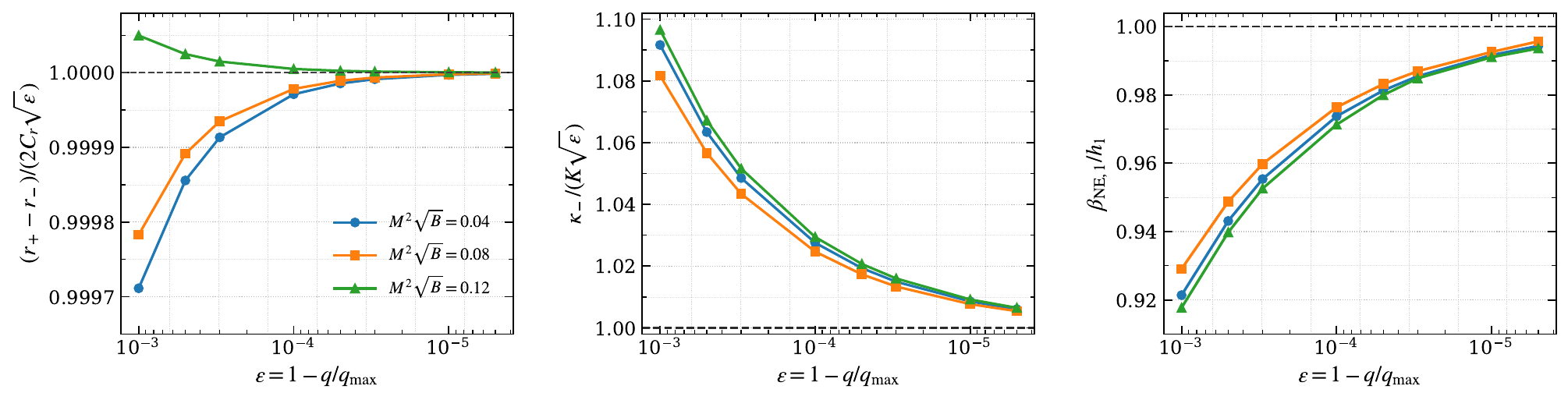}
 \caption{Near-extremal scaling at fixed $B$ as $\epsilon=1-q/\qmax\to0$; the horizontal axis is logarithmic.  Left: horizon splitting normalized by its leading prediction, $(r_+-r_-)/(2C_r\sqrt\epsilon)$.  Center: Cauchy-horizon surface gravity normalized by its leading prediction, $\kappam/(K\sqrt\epsilon)$.  Right: $\ell=1$ NE damping ratio normalized by the conformal weight, $\beta_{{\rm NE},1}/h_1$.  In all panels the black dashed line marks unity, the analytic cold-limit prediction.}
 \label{fig:NEscaling}
\end{figure*}

At extremality on the cold branch, the event-horizon temperature vanishes, $T_+=\kappap/(2\pi)\to0$, while the cosmological horizon remains generically nondegenerate.  The quadratic part of the lapse then defines an $\mathrm{AdS}_2\times S^2$ throat \cite{Romans1992,CastroMarianiToldo2023} with
\begin{equation}
 L_2^2=\frac{2r_e^2}{D_e}.
 \label{eq:L2}
\end{equation}
For the massless scalar, spherical reduction gives the effective $\mathrm{AdS}_2$ mass $m_{\rm eff}^2=\ell(\ell+1)/r_e^2$.  The corresponding conformal weight,
\begin{equation}
 h_\ell(B)=\frac12+\sqrt{\frac14+\frac{2\ell(\ell+1)}{D_e}},
 \label{eq:hell}
\end{equation}
is the decaying radial exponent at the $\mathrm{AdS}_2$ boundary and satisfies $h_\ell(h_\ell-1)=m_{\rm eff}^2L_2^2$.  The near-horizon spectrum therefore obeys the scaled cold-limit relation
\begin{equation}
 \frac{\omega_{{\rm NE},\ell n}}{\kappam}
 \longrightarrow -i\bigl[n+h_\ell(B)\bigr],
 \label{eq:NEanalytic}
\end{equation}
where $n=0,1,2,\ldots$ is the overtone number, consistent with the general near-horizon description of zero-damped modes in nearly extremal cosmological black holes \cite{Joykutty2022,Davey2024,HintzNE2025}.  Since $\kappap/\kappam\to1$, either horizon surface gravity gives the same leading scaled limit.  Therefore
\begin{equation}
 \beta_{{\rm NE},\ell n}\longrightarrow n+h_\ell(B).
 \label{eq:betaNEanalytic}
\end{equation}
Because $r_e(B)$ and $D_e(B)$ are fixed analytically by the cold-boundary parametrization, Eqs.~\eqref{eq:hell}--\eqref{eq:betaNEanalytic} give the CDF-specific scaled NE spectrum explicitly along the cold branch.  For $\ell=0$, $h_0=1$ and the fundamental mode satisfies $\beta_{\rm NE}\to1$, recovering the massless CDF near-extremal result of Ref.~\cite{Becar2024}.  For higher multipoles, the scaled damping retains explicit dependence on the extremal CDF geometry through $h_\ell(B)$.  The $\ell=0$ limit is recovered in the global numerical results of Fig.~\ref{fig:beta}, while Fig.~\ref{fig:NEscaling} directly tests the first geometry-dependent case, $\ell=1$.

Figure~\ref{fig:NEscaling} directly tests these three near-extremal predictions against the numerical solutions.  The left and central panels verify the horizon-splitting and surface-gravity coefficients in Eqs.~\eqref{eq:rpmexpansion} and \eqref{eq:kappaexpansion}, respectively.  The right panel tests the CDF-dependent conformal weight through the $\ell=1$ NE mode.

For SCC, the analytic NE result fixes the cold-endpoint behavior of the spectral gap rather than the location of the SCC threshold.  Since the dS and PS damping rates remain finite while $\kappam\to0$, their individual ratios $-\operatorname{Im}\omega/\kappam$ diverge toward extremality.  The NE damping instead scales with $\kappam$, giving $\beta_{\rm NE}\to1$; sufficiently close to the cold endpoint, the NE family therefore controls the lower envelope.  The $\beta=1/2$ threshold can nevertheless be crossed earlier while a dS or PS mode remains dominant.  The global competition among these families is determined in Sec.~\ref{sec:results}.

\section{Numerical methods and results}
\label{sec:results}

\subsection{Numerical methods and cross-checks}
\label{sec:num_methods}

We compute the QNM frequencies using a Chebyshev pseudospectral method, following the standard spectral strategy for black-hole QNMs \cite{Jansen2017}. We map the exterior interval $r\in[r_+,r_c]$ to $z\in[-1,1]$ by
\begin{equation}
    r=\frac{r_c-r_+}{2}z+\frac{r_c+r_+}{2}
    \label{eq:map}
\end{equation}
and factor out the known QNM behavior at both endpoints,
\begin{equation}
    \psi(r)=
    (1+z)^{-i\omega/(2\kappap)}
    (1-z)^{-i\omega/(2\kappac)}y(z),
    \label{eq:factor}
\end{equation}
where constant factors relating $1\pm z$ to the corresponding horizon distances are absorbed into the overall normalization.  The reduced radial function $y(z)$ is regular on the compact interval and is expanded in Chebyshev polynomials of the first kind,
\begin{equation}
    y(z)=\sum_{j=0}^{N}a_j T_j(z),
    \label{eq:chebexp}
\end{equation}
where $T_j(z)=\cos[j\arccos(z)]$ and $a_j$ are the spectral expansion coefficients.  We collocate the radial equation at the Chebyshev--Gauss points $z_i=\cos[(i+1/2)\pi/(N+1)]$, $i=0,\ldots,N$. Substitution into Eq.~\eqref{eq:radial} produces a quadratic matrix eigenvalue problem
\begin{equation}
    \left(M_0+\omega M_1+\omega^2M_2\right)\bm a=0,
    \label{eq:quadratic}
\end{equation}
where $\bm a=(a_0,\ldots,a_N)^T$.  The matrices $M_0$, $M_1$, and $M_2$ are the $(N+1)\times(N+1)$ collocation matrices assembled from the Chebyshev differentiation matrices and the background coefficients; they collect the terms proportional to $\omega^0$, $\omega^1$, and $\omega^2$, respectively.  The quadratic problem is linearized to a $2(N+1)$-dimensional generalized eigenvalue problem.  Physical roots are identified by stability under increasing spectral order and by continuation in the black-hole parameters.  Generic PS and dS branches are followed in $\omega$, while the NE sector is tracked using the ratio $\omega/\kappam$.  Successive spectral orders are compared for the near-extremal checks, with higher orders employed as extremality is approached.

Representative modes are independently checked by direct integration \cite{KonoplyaZhidenko2011}. Local series solutions satisfying the two QNM boundary conditions are integrated from the event and cosmological horizons to an interior matching point, where equality of logarithmic derivatives determines the frequency. The PS sector is further checked with the leading Wentzel--Kramers--Brillouin (WKB) barrier condition \cite{SchutzWill1985,Konoplya2003}. For the large-$\ell$ candidate entering the SCC scan we use the eikonal WKB damping rate; in the finite-$\ell$ plots this limit is represented numerically by $\ell=500$.

Table~\ref{tab:validation} gives representative damping ratios at $q/\qmax=0.99$ and $0.995$. At low and moderate multipoles, the pseudospectral and direct-integration values agree closely. In the PS sector, the pseudospectral and WKB results at $\ell=20$ differ by less than $0.02\%$ for all entries shown, demonstrating convergence toward the eikonal damping rate. At a few of the highest-$\ell$ near-extremal points, the direct-integration values show larger deviations, whereas the pseudospectral and WKB results remain mutually consistent.

\begin{table*}[!t]
\centering
\footnotesize
\caption{Representative lowest-lying nontrivial scalar modes, characterized by $\beta_\ell=-\operatorname{Im}\omega_\ell/\kappam$, at two near-extremal ratios. Pseudospectral values are checked against direct integration; the PS sector is additionally compared with WKB. A dash indicates that the corresponding method was not used for that entry.}
\label{tab:validation}
\setlength{\tabcolsep}{4.2pt}
\renewcommand{\arraystretch}{0.94}
\begin{tabular}{c c l c c c c c}
\toprule
$q/\qmax$ & $M^2\sqrt B$ & Method & $\ell=0$ & $\ell=1$ & $\ell=2$ & $\ell=10$ & $\ell=20$ \\
\midrule
\multirow{9}{*}{0.99}
& \multirow{3}{*}{0.04} & Pseudospectral      & 0.727730 & 0.560687 & 0.541204 & 0.529699 & 0.529109 \\
&                         & Direct integration & 0.727664 & 0.560716 & 0.541226 & 0.529647 & 0.529121 \\
&                         & WKB                & --       & --       & --       & 0.529561 & 0.529202 \\
\cmidrule(lr){2-8}
& \multirow{3}{*}{0.08} & Pseudospectral      & 0.831065 & 0.669297 & 0.648731 & 0.637833 & 0.637319 \\
&                         & Direct integration & 0.831319 & 0.669297 & 0.648827 & 0.637833 & 0.636878 \\
&                         & WKB                & --       & --       & --       & 0.637779 & 0.637304 \\
\cmidrule(lr){2-8}
& \multirow{3}{*}{0.12} & Pseudospectral      & 0.709312 & 0.506815 & 0.503708 & 0.502311 & 0.502252 \\
&                         & Direct integration & 0.709363 & 0.506815 & 0.503708 & 0.502311 & 0.502914 \\
&                         & WKB                & --       & --       & --       & 0.502252 & 0.502236 \\
\midrule
\multirow{9}{*}{0.995}
& \multirow{3}{*}{0.04} & Pseudospectral      & 0.865631 & 0.858776 & 0.828385 & 0.810649 & 0.809974 \\
&                         & Direct integration & 0.865631 & 0.858788 & 0.828354 & 0.811506 & 0.805723 \\
&                         & WKB                & --       & --       & --       & 0.810470 & 0.809903 \\
\cmidrule(lr){2-8}
& \multirow{3}{*}{0.08} & Pseudospectral      & 0.857464 & 1.016186 & 0.984855 & 0.968173 & 0.967383 \\
&                         & Direct integration & 0.857464 & 1.016212 & 0.984835 & 0.967178 & 0.968738 \\
&                         & WKB                & --       & --       & --       & 0.968105 & 0.967364 \\
\cmidrule(lr){2-8}
& \multirow{3}{*}{0.12} & Pseudospectral      & 0.754666 & 0.778234 & 0.773697 & 0.771645 & 0.771558 \\
&                         & Direct integration & 0.754666 & 0.778234 & 0.773697 & 0.771645 & 0.771418 \\
&                         & WKB                & --       & --       & --       & 0.771585 & 0.771543 \\
\bottomrule
\end{tabular}
\end{table*}

\subsection{QNM spectral overview}

\begin{figure*}[!t]
    \centering
    \includegraphics[width=0.99\textwidth]{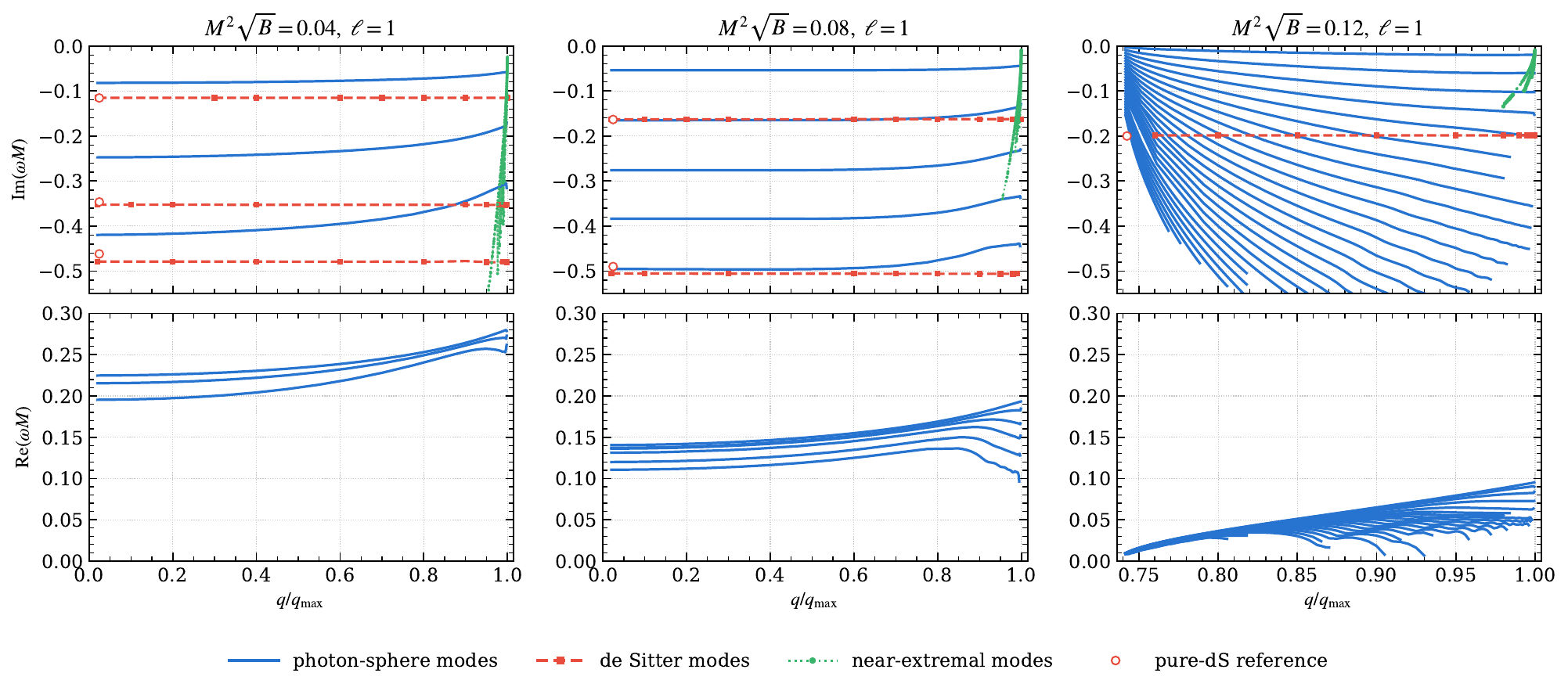}
    \caption{Overview of the $\ell=1$ scalar QNM spectrum as a function of $q/\qmax$ for $M^2\sqrt B=0.04$, $0.08$, and $0.12$ (left to right).  The upper and lower rows show $\operatorname{Im}(\omega M)$ and $\operatorname{Re}(\omega M)$, respectively.  Blue curves are PS overtones, red dashed curves are dS modes, and green markers show NE modes near extremality; open red circles indicate the corresponding pure-de Sitter reference frequencies.  Only the branches reliably identified in our numerical scan are shown; the figure is not intended as an exhaustive enumeration of all QNM overtones.}
    \label{fig:spectrum}
\end{figure*}

The scalar spectrum contains the three families identified for the CDF geometry in Ref.~\cite{Becar2024}: complex PS modes associated with unstable null trapping, purely or predominantly imaginary dS modes tied to the asymptotic de Sitter region, and purely imaginary NE modes that become long lived as $r_-\to r_+$.  Figure~\ref{fig:spectrum} gives an overview of the $\ell=1$ spectrum for three representative values of $M^2\sqrt B$.  Several overtones are retained to display the spectral structure.  The figure is intended as a representative spectral overview rather than a complete catalogue of QNM overtones.

For the dS family, the pure-de Sitter limit provides a natural reference scale. With $\Lambda_{\rm eff}=\sqrt B$, the fundamental massless $\ell=1$ frequency of empty de Sitter space is
\begin{equation}
    \omega_{\rm dS}^{(0)}=-i\sqrt{\frac{\sqrt B}{3}}.
    \label{eq:dSestimate}
\end{equation}
The CDF geometry shifts the dS frequencies away from their pure-de Sitter values, but Eq.~\eqref{eq:dSestimate} sets their characteristic damping scale. As $M^2\sqrt B$ increases, the dS branches move to larger damping rates, while the relative position of the PS branches changes accordingly; the same trend was found in the mode-dominance analysis of Ref.~\cite{Becar2024}.

\subsection{Spectral-gap competition and SCC-violation region}

For SCC, the relevant quantity is not the full spectrum but its slowest-decaying member. Following the family-based strategy used for RNdS and PFDM black holes \cite{Cardoso2018,Jiang2024}, we define
\begin{equation}
    \beta_\ell(B,q)=\min_n\left[-\frac{\operatorname{Im}\omega_{\ell n}(B,q)}{\kappam(B,q)}\right],
    \label{eq:betaell}
\end{equation}
and take the lower envelope
\begin{equation}
    \betamin(B,q)=\min_\ell\beta_\ell(B,q).
    \label{eq:betamin}
\end{equation}
In constructing this lower envelope, we retain only branches that pass the convergence and continuation checks described in Sec.~\ref{sec:num_methods} and Appendix~\ref{app:numerics}.  The relevant candidates are the lowest dS mode in the low-multipole sector, the PS mode approaching the large-$\ell$ eikonal limit, and the lowest NE branch. Finite-$\ell$ and WKB comparisons in Table~\ref{tab:validation} control the approach of the PS damping rate to its eikonal value.

The NE family provides the analytic check derived in Sec.~\ref{sec:cold}. The fundamental $\ell=n=0$ member tends to $\beta_{\rm NE}=1$ and prevents the spectral-gap ratio from diverging even though the dS and PS ratios would grow as $\kappam\to0$ if considered separately. The $\ell=1$ branches shown in Fig.~\ref{fig:spectrum} instead approach the $B$-dependent values in Eq.~\eqref{eq:hell}.

\begin{figure*}[!t]
    \centering
    \includegraphics[width=0.99\textwidth]{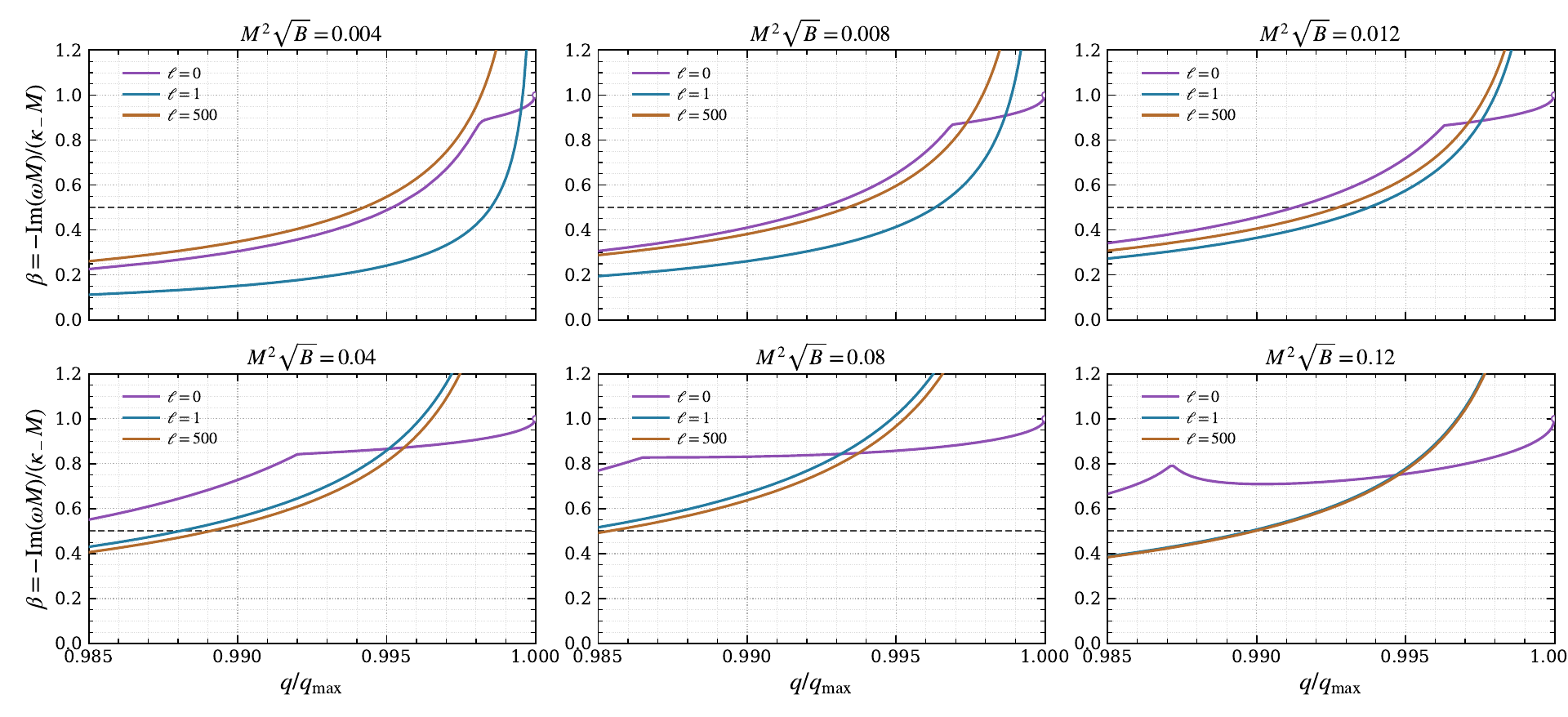}
    \caption{Lowest-lying damping ratios $\beta_\ell=-\operatorname{Im}(\omega M)/(\kappam M)$ as functions of $q/\qmax$ for six representative values of $M^2\sqrt B$.  Purple and blue curves show the $\ell=0$ and $\ell=1$ candidates, and the brown curve denotes the eikonal PS contribution, represented numerically by $\ell=500$.  The horizontal dashed line marks the Christodoulou threshold $\beta=1/2$; the lower envelope of these candidates determines the spectral gap used below.  The horizontal range extends to $q/\qmax=1$, making the fundamental NE limit $\beta\to1$ explicit.}
    \label{fig:beta}
\end{figure*}

Figure~\ref{fig:beta} displays the lowest-lying ratios for $M^2\sqrt B=0.004, 0.008, 0.012, 0.04, 0.08,$ and $0.12$, using the $\ell=0$, $\ell=1$, and large-$\ell$ PS candidates. At each value of $q/\qmax$, the spectral gap is determined by the lowest of these curves. For the smaller values of $M^2\sqrt B$, the $\ell=1$ dS branch forms the lower envelope at the $\beta=1/2$ crossing and therefore determines $q_{\rm crit}$. At intermediate $M^2\sqrt B$, the large-$\ell$ PS branch lies below the dS branch near the crossing and sets the threshold instead; a second switch near the ultracold endpoint is resolved below. Closer to extremality, the fundamental NE branch becomes the lowest mode and approaches $\beta=1$ in accordance with Eq.~\eqref{eq:betaNEanalytic}. The SCC threshold is crossed before the extremal boundary.  Thus, at fixed $B$, the potential-violation region is the finite interval
\begin{equation}
    q_{\rm crit}(B)<q<\qmax(B),
    \label{eq:violation_interval}
\end{equation}
throughout which $\betamin>1/2$.

For each $B$, the SCC threshold $q_{\rm crit}(B)$ is defined by
\begin{equation}
    \betamin\bigl(B,q_{\rm crit}(B)\bigr)=\frac12.
    \label{eq:qcrit}
\end{equation}
We normalize the CDF inhomogeneity parameter by its exact extremal value,
\begin{equation}
    \qhat\equiv\frac{q}{\qmax(B)},
    \qquad
    \Delta\qhat\equiv1-\qhat_{\rm crit},
    \label{eq:width}
\end{equation}
so that $\qhat=1$ is extremality for every $B$ and $\Delta\qhat$ measures the normalized width of the violation region directly.  Unless stated otherwise, ``width'' below refers to this normalized width, not to the interval in the unnormalized parameter $q$.

\begin{figure*}[!t]
    \centering
    \includegraphics[width=0.485\textwidth]{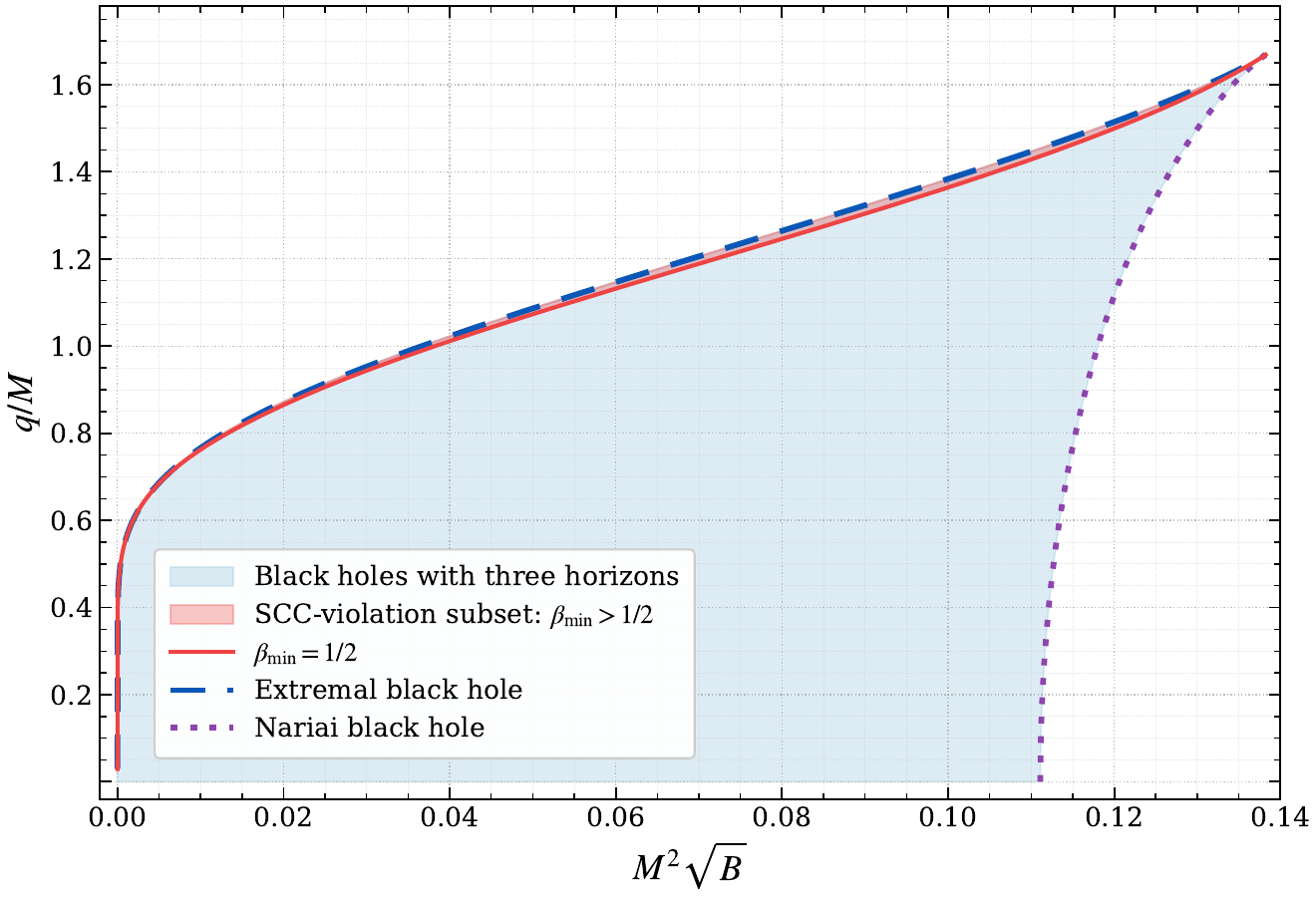}\hfill
    \includegraphics[width=0.485\textwidth]{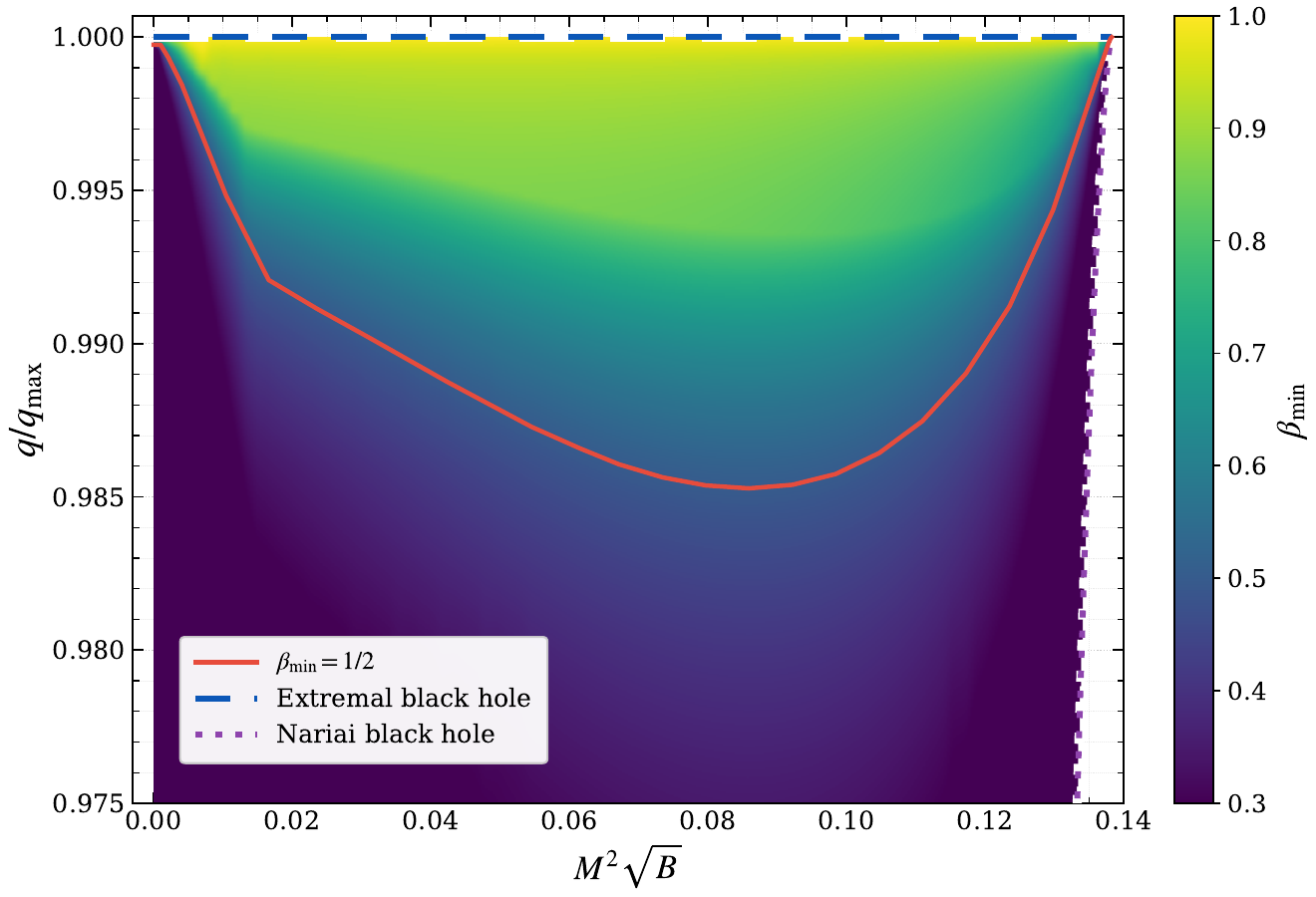}
    \caption{Massless-scalar SCC phase diagram for $M=1$.  (a) Physical $(M^2\sqrt B,q/M)$ parameter plane.  The light-blue region is the three-horizon domain, and the semi-transparent red strip is the subset satisfying $\betamin>1/2$, bounded below by the solid red curve $\betamin=1/2$ and above by the dashed extremal boundary.  (b) Zoomed normalized $(M^2\sqrt B,q/\qmax)$ plane.  The color map shows $\betamin$, the solid red curve again marks $\betamin=1/2$, the dashed blue line is the extremal boundary $q/\qmax=1$, and the dotted purple curve is the Nariai boundary.  This panel resolves the nonmonotonic width of the SCC-violation region near extremality.}
    \label{fig:violation}
\end{figure*}

Figure~\ref{fig:violation} summarizes the SCC phase diagram in physical and normalized coordinates.  The potential-violation region remains adjacent to the extremal curve throughout the three-horizon domain, but its normalized width is nonmonotonic, narrowing toward both small $M^2\sqrt B$ and the triple-horizon endpoint.

\begin{figure}[!t]
 \centering
 \includegraphics[width=\columnwidth]{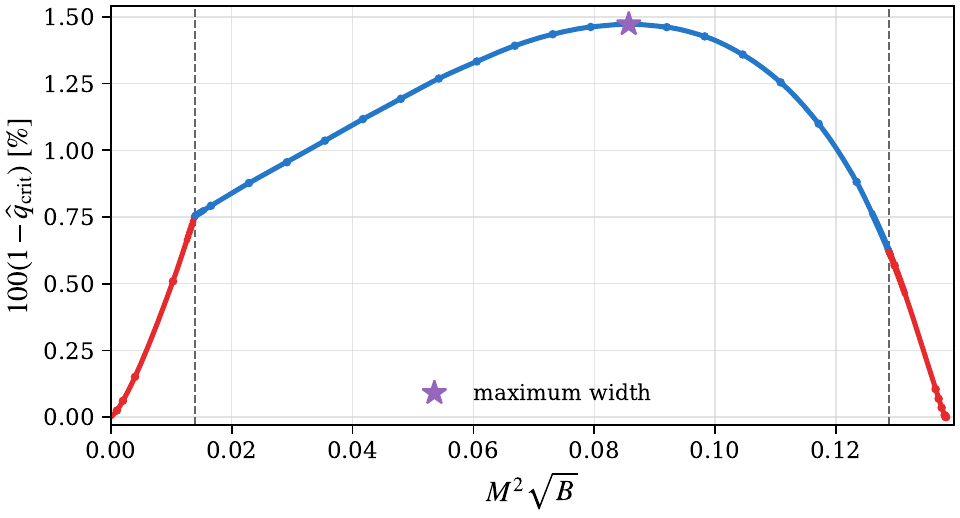}
 \caption{Normalized width of the SCC-violation region, $\Delta\qhat=1-\qhat_{\rm crit}$, resolved by the mode family controlling the threshold.  Red and blue points identify dS- and PS-controlled portions of the threshold, respectively.  The vertical dashed lines mark the two controller exchanges, and the star denotes the maximum normalized width.}
 \label{fig:controller}
\end{figure}

To identify which QNM family sets the SCC threshold, Fig.~\ref{fig:controller} shows the normalized width $\Delta\qhat=1-\qhat_{\rm crit}$ with the controlling family encoded along the curve.  The dS branch controls the threshold at small $M^2\sqrt B$, the eikonal PS branch takes over near $M^2\sqrt B\simeq0.0139$, and the $\ell=1$ dS branch regains control close to the ultracold endpoint near $M^2\sqrt B\simeq0.1288$.  The maximum occurs within the PS-controlled interval, at $M^2\sqrt B\simeq0.0857$, where $\qhat_{\rm crit}\simeq0.9853$ and $\Delta\qhat_{\rm max}\simeq0.0147$, or about $1.47\%$.  Local branch-resolved refinement around both exchanges confirms the controller assignment; the corresponding convergence checks are summarized in Appendix~\ref{app:numerics}.

\subsection{Comparison with the PFDM case}
\label{sec:PFDMcomparison}

For comparison, we consider the perfect-fluid-dark-matter (PFDM) black hole of Ref.~\cite{Jiang2024}, which likewise exhibits a narrow near-extremal SCC-violation region.  In that model the normalized violation region first broadens and then narrows as the cosmological constant increases, while the published QNM data associate the broadening and narrowing segments with dS- and PS-controlled decay, respectively.  We independently reproduced the QNM benchmarks and confirmed that the dS--PS exchange and the maximum normalized width fall in the same narrow parameter interval at the resolution of our check.  The cold-limit analysis gives the same qualitative picture: the PFDM PS-width estimate decreases monotonically along the physical cold branch, so once the PS family controls the threshold, the normalized width is already decreasing.  The numerical and cold-limit details are collected in Appendix~\ref{app:PFDM}.

The CDF phase boundary shows a different pattern.  Its first dS--PS exchange occurs near $M^2\sqrt B\simeq0.0139$, whereas the maximum normalized width occurs much later, at $M^2\sqrt B\simeq0.0857$.  Thus the PS-controlled CDF segment continues to broaden after the controller exchange, reaches a maximum, and only then narrows.  The cold scaling derived in Sec.~\ref{sec:cold} provides an analytic estimate of this behavior.

At fixed $B$ write $q=q_e(1-\epsilon)$ with $q_e=\qmax(B)$, as in Eq.~\eqref{eq:epsilon}.  For either finite-damping family $X\in\{\mathrm{dS},\mathrm{PS}\}$, let $\epsilon_{{\rm crit},X}=1-\qhat_{{\rm crit},X}=\Delta\qhat_X$ denote its branch-specific SCC threshold.  Its damping rate tends to a finite cold value, $\alpha_X=\alpha_{e,X}+O(\epsilon)$, while Eq.~\eqref{eq:kappaexpansion} gives $\kappam=K(B)\sqrt{\epsilon}+O(\epsilon)$.  Imposing $\beta_X=\alpha_X/\kappam=1/2$ therefore gives
\begin{equation}
 \Delta\qhat_X(B)\simeq
 \left[\frac{2\alpha_{e,X}(B)}{K(B)}\right]^2.
 \label{eq:branchwidth}
\end{equation}
For the eikonal PS family the cold damping rate has a direct geometric approximation.  If $r_{e,{\rm ph}}$ denotes the unstable circular null orbit of the extremal CDF background, then \cite{CardosoGeodesic2009}
\begin{align}
 \alpha_{e,{\rm PS}}&\simeq\frac{\lambda_{e,{\rm ph}}}{2},\\
 \lambda_{e,{\rm ph}}^2&=
 \left.\frac{f\bigl(2f-r^2f''\bigr)}{2r^2}\right|_{r=r_{e,{\rm ph}},\,q=q_e}.
 \label{eq:PScold}
\end{align}
Combining these relations yields the leading geometric estimate
\begin{equation}
 \Delta\qhat_{\rm PS}(B)\simeq
 \left[\frac{\lambda_{e,{\rm ph}}(B)}{K(B)}\right]^2.
 \label{eq:PSwidthcold}
\end{equation}
Accordingly, the extremum of the leading PS-controlled width is set by the stationary point of $\lambda_{e,{\rm ph}}/K$ (equivalently $\alpha_{e,{\rm PS}}/K$), whereas a dS--PS controller exchange is set by equality of the two damping rates.  Evaluating Eq.~\eqref{eq:PScold} on the exact cold boundary predicts the maximum of $\lambda_{e,{\rm ph}}/K$ at $M^2\sqrt B\simeq0.0853$, close to the full numerical maximum at $0.0857$.  Equation~\eqref{eq:branchwidth} is less accurate for the absolute normalized width because the threshold occurs at $\Delta\qhat\sim10^{-2}$ rather than in the asymptotically infinitesimal regime, but it captures the nonmonotonic trend and the location of the maximum.  Detailed finite-$\qhat$ WKB thresholds and cold-limit comparisons are given in Appendix~\ref{app:numerics}.

The two geometries differ in how the parameter controlling the asymptotic de Sitter scale enters the strong-field geometry.  In PFDM, the cosmological constant $\Lambda$ and the dark-matter strength $b$ are independent parameters.  In the CDF solution, by contrast, $B$ is not merely an asymptotic cosmological scale: it also enters the full radial density profile.  Varying $B$ therefore changes the effective de Sitter curvature together with the extremal geometry, the coefficient $K(B)$, and the photon potential.  This coupled dependence underlies the separation between the first dS--PS exchange and the subsequent width maximum in the CDF case.
\FloatBarrier

\section{Conclusions}
\label{sec:conclusion}

We have investigated strong cosmic censorship for electrically neutral black holes immersed in a Chaplygin-like dark fluid using the massless-scalar QNM spectrum and analytic control of the extremal geometry.  The exact parametrization of the Nariai and cold degenerate-horizon branches leads to the near-extremal horizon splitting and Cauchy-horizon surface-gravity scalings in Eqs.~\eqref{eq:rpmexpansion} and \eqref{eq:kappaexpansion}, which are verified numerically in Fig.~\ref{fig:NEscaling}.  The associated near-horizon analysis gives Eqs.~\eqref{eq:NEanalytic} and \eqref{eq:betaNEanalytic}: the fundamental $\ell=n=0$ mode approaches $\beta_{\rm NE}=1$, while higher multipoles retain explicit dependence on the cold CDF geometry through $h_\ell(B)$.

The global spectral gap is set by the competition among dS, PS, and NE families.  The Christodoulou threshold is crossed before extremality, producing the narrow potential-violation region shown in Fig.~\ref{fig:violation}.  As resolved in Fig.~\ref{fig:controller}, the threshold controller changes from dS to PS near $M^2\sqrt B\simeq0.0139$ and back to dS near $0.1288$, whereas the normalized width reaches its maximum at the distinct value $M^2\sqrt B\simeq0.0857$.  The PS-controlled segment is captured by the cold-limit estimate in Eq.~\eqref{eq:PSwidthcold}, $\Delta\qhat_{\rm PS}\simeq(\lambda_{e,{\rm ph}}/K)^2$, whose stationary point at $M^2\sqrt B\simeq0.0853$ closely reproduces the numerical maximum.  Sufficiently close to the cold endpoint, the NE family instead controls the limiting spectral behavior.  In the PFDM comparison, by contrast, the PS-controlled cold-width estimate is already decreasing when the PS family becomes the threshold controller.

These conclusions apply to a linear massless scalar and the Christodoulou $H^1$ criterion.  Coupled metric--CDF perturbations and nonlinear backreaction may modify the fate of the Cauchy horizon.  The ultracold triple root also requires a separate scaling analysis beyond the ordinary cold expansion considered here.

\appendix
\section{Exact parametrization of the degenerate-horizon boundaries}
\label{app:horizon}

This appendix derives the exact parametric form of the CDF degenerate-horizon boundaries quoted in Sec.~\ref{sec:geometry} and clarifies its relation to the horizon analysis of Ref.~\cite{Fontana2026}. The result concerns the \emph{boundary} configurations in which two or three horizons coincide; generic nondegenerate roots of $f(r)=0$ still require numerical solution.

\subsection{Reduction of the degenerate-horizon conditions}

For the anisotropic CDF source one may write
\begin{equation}
    T^{\mu}{}_{\nu}=\operatorname{diag}(-\rho,-\rho,p_t,p_t),
    \qquad
    p_t=\frac{\rho}{2}-\frac{3B}{2\rho},
    \label{eq:appT}
\end{equation}
where the second relation follows from the angularly averaged equation of state $p=-B/\rho$. The independent Einstein equations are
\begin{align}
    \frac{f+rf'-1}{r^2}&=-\rho,\label{eq:appE1}\\
    \frac{2f'+rf''}{2r}&=\frac{\rho}{2}-\frac{3B}{2\rho}.\label{eq:appE2}
\end{align}
Let $r_d$ be a degenerate positive root,
\begin{equation}
    f(r_d)=0,\qquad f'(r_d)=0.
    \label{eq:appdeg}
\end{equation}
Equation~\eqref{eq:appE1} then immediately yields
\begin{equation}
    \rho(r_d)=\frac{1}{r_d^2}.
    \label{eq:apprho}
\end{equation}
Denoting the parameter values on the degenerate boundary by $q_d$ and $B_d$, we introduce
\begin{equation}
    u_d\equiv\frac{q_d}{r_d},\qquad 0\le u_d<1,
    \label{eq:appu}
\end{equation}
Equation~\eqref{eq:rho} then gives
\begin{equation}
    B_d=\frac{1-u_d^2}{r_d^4}.
    \label{eq:appB}
\end{equation}
The restriction $u_d<1$ follows directly from $B_d>0$.  This auxiliary variable is used only to reduce the degenerate-horizon conditions; the main text uses $x=\operatorname{arctanh}u_d$ as the boundary parameter.  Moreover,
\begin{equation}
    \operatorname{arcsinh}\!\left(\frac{u_d}{\sqrt{1-u_d^2}}\right)
    =\operatorname{arctanh}u_d,
    \label{eq:appidentity}
\end{equation}
so the remaining condition $f(r_d)=0$ reduces to
\begin{equation}
    2-\frac{6M}{r_d}+u_d\operatorname{arctanh}u_d=0.
    \label{eq:appfred}
\end{equation}
Hence the entire degenerate boundary is
\begin{align}
    r_d(u_d)&=\frac{6M}{2+u_d\operatorname{arctanh}u_d},\label{eq:apprdu}\\
    q_d(u_d)&=u_d\,r_d(u_d),\label{eq:appqdu}\\
    B_d(u_d)&=\frac{1-u_d^2}{r_d(u_d)^4}.\label{eq:appBdu}
\end{align}
Introducing the main-text parameter $x=\operatorname{arctanh}u_d$, or $u_d=\tanh x$, removes the endpoint singularity at $u_d\to1$ and gives Eqs.~\eqref{eq:rdx}--\eqref{eq:Bdx}.

\subsection{Nariai, extremal, and triple-horizon branches}

The sign of $f\,^{\prime\prime}(r_d)$ gives an analytic classification of the two degenerate branches. Evaluating Eq.~\eqref{eq:appE2} on a degenerate horizon and using Eqs.~\eqref{eq:apprho} and \eqref{eq:appB} gives
\begin{equation}
    f''(r_d)=\frac{3u_d^2-2}{r_d^2}.
    \label{eq:appfpp}
\end{equation}
Thus $u_d^2<2/3$ gives $f''(r_d)<0$, so the double root is a local maximum of $f$ and corresponds to the merger $r_+=r_c$ (Nariai). For $u_d^2>2/3$, $f''(r_d)>0$ and the double root is a local minimum, corresponding to $r_-=r_+$ (extremal). At the triple-horizon point,
\begin{equation}
    u_{d,\star}=\sqrt{\frac23},
    \qquad
    x_\star=\operatorname{arctanh}\sqrt{\frac23},
    \label{eq:appuc}
\end{equation}
we have
\begin{equation}
    f(r_\star)=f'(r_\star)=f''(r_\star)=0.
    \label{eq:apptriple}
\end{equation}
A further differentiation gives $f^{(3)}(r_\star)=-4/r_\star^3\neq0$, confirming that this is a genuine third-order zero. For $M=1$, Eqs.~\eqref{eq:apprdu}--\eqref{eq:appBdu} give
\begin{equation}
    r_\star\simeq2.04368,\qquad
    q_\star\simeq1.66866,\qquad
    \sqrt{B_\star}\simeq0.138234,
    \label{eq:appstarnumbers}
\end{equation}
together with
\begin{equation}
    B_\star q_\star^4=\frac{4}{27}.
    \label{eq:appcritical}
\end{equation}
The point is the ultracold limit in the usual de Sitter black-hole terminology.
The other endpoint provides a useful check: at $x=0$ one has $q=0$, $r_d=3M$, and $M^4B=1/81$, precisely the Schwarzschild--de Sitter Nariai point. In the opposite limit $x\to\infty$, the extremal boundary approaches $(B,q)\to(0,0)$ while $r_d\to0$; the parameter-space origin itself is the Schwarzschild solution and should not be interpreted as a finite-radius extremal black hole.

\subsection{Relation to previous horizon analysis}

Ref.~\cite{Fontana2026} analyzed the global shape of the same CDF metric, proved that $f(r)$ can have at most three positive roots, and derived the critical relation terminating the three-horizon domain. Their variables are related to those used here by
\begin{equation}
    B_{\rm F}=\frac{B}{9},\qquad Q=\frac{q}{3}.
    \label{eq:appmapping}
\end{equation}
Accordingly, Eq.~\eqref{eq:appcritical} becomes
\begin{equation}
    B_{\rm F}Q^4=\frac{4}{3^9},
    \label{eq:appFontanacrit}
\end{equation}
and $q_\star/3\simeq0.556219M$, reproducing their global critical values.

Ref.~\cite{Fontana2026} establishes the global horizon multiplicity and causal structure and obtains the two threshold curves numerically once one parameter is fixed. Equations~\eqref{eq:apprdu}--\eqref{eq:appBdu} provide a unified exact parametric representation of the same Nariai and extremal boundaries. In particular, for a prescribed $B$ the extremal boundary value used as the reference in the SCC analysis is obtained by solving the single equation $B_d(x)=B$ on the monotonic extremal branch $x>x_\star$ and setting
\begin{equation}
    \qmax(B)=q_d(x).
    \label{eq:appqmax}
\end{equation}
This replaces a simultaneous numerical solution of $f=f'=0$ by a one-dimensional inversion and is especially convenient for near-extremal scans.

\section{Numerical branch tracking and phase-boundary validation}
\label{app:numerics}

This appendix documents the numerical checks underlying the spectral-branch assignment and the SCC phase boundary.  The phase boundary is evaluated on a dense set of $B$ values, supplemented by logarithmic sampling at small $B$ and targeted meshes around both controller exchanges.  For each $B$, the first crossing of $\betamin=1/2$ is bracketed from the extremal side and refined by bisection and local interpolation.  The crossing and the neighboring controller values are then recomputed at higher spectral order.

For low multipoles the generalized eigenvalue spectra at successive Chebyshev orders are compared by mutual nearest-neighbor matching.  Roots with positive damping, the static $\ell=0$ solution, and roots not reproduced at the next order are discarded.  Branch identity is then assigned by parameter continuation, not by sorting the roots independently at every point.  The eikonal PS candidate is evaluated from the barrier maximum and checked against $\ell=10$ and $20$ pseudospectral data.  The dS branches are additionally identified by continuation from their pure-de Sitter reference frequencies on the imaginary axis.

The NE calculation uses the scaled variable $\omega/\kappam$, which keeps the relevant eigenvalues finite as $\kappam\to0$.  The $\ell=1$ sequence in Fig.~\ref{fig:NEscaling} is followed by continuation from the cold end and checked against adjacent spectral orders; the production orders are increased for the smaller-$B$ backgrounds, where the nearly coincident horizons are numerically more demanding.

Table~\ref{tab:switchaudit} summarizes the convergence check near both exchanges and at the maximum-width point.  The relative ordering of the dS and PS candidates is stable as the spectral order is increased, and the dS candidate remains above the PS controller at the maximum-width point.

\begin{table}[!htbp]
\caption{Convergence check for the mode controlling the SCC threshold. Values are evaluated at Chebyshev order $N=60$ and at the refined $\qhat_{\rm crit}$.}
\label{tab:switchaudit}
\begin{ruledtabular}
\begin{tabular}{c c c c}
$M^2\sqrt B$ & $\beta_{\rm dS}$ & $\beta_{\rm PS}$ & controller\\
\hline
0.013919411 & 0.500000009 & 0.500023565 & dS\\
0.013920488 & 0.500001942 & 0.499999694 & PS\\
0.085721914 & 0.522867013 & 0.500000046 & PS\\
0.128807997 & 0.500000244 & 0.500000172 & PS\\
0.128808391 & 0.499999999 & 0.500000007 & dS\\
\end{tabular}
\end{ruledtabular}
\end{table}

For Fig.~\ref{fig:controller}, the branch-specific dS and PS thresholds were additionally sampled on dense local meshes around both exchanges.  These local branch-resolved scans reproduce the two controller-exchange locations quoted in the main text.

In the PS-controlled interval, we compare the cold photon-sphere estimate in Eq.~\eqref{eq:PSwidthcold} with the full finite-$\qhat$ WKB condition $\beta_{\rm PS}=1/2$.  The cold estimate captures the shape and the maximum position but overestimates the absolute value of the normalized width, as expected for a leading expansion evaluated at $\Delta\qhat\sim10^{-2}$.  Representative values are listed in Table~\ref{tab:PSwidth}.
\begin{table}[!htbp]
\caption{PS-controlled normalized SCC-violation width from the finite-$\qhat$ WKB threshold and from the cold estimate in Eq.~\eqref{eq:PSwidthcold}.}
\label{tab:PSwidth}
\begin{ruledtabular}
\begin{tabular}{c c c}
$M^2\sqrt B$ & $\Delta\qhat_{\rm PS}$ (WKB) & cold estimate\\
\hline
0.0400 & 0.01092 & 0.01912\\
0.0800 & 0.01464 & 0.02624\\
0.0857 & 0.01472 & 0.02638\\
0.1200 & 0.01007 & 0.01772\\
\end{tabular}
\end{ruledtabular}
\end{table}

\section{Near-extremal comparison with the PFDM geometry}
\label{app:PFDM}

For comparison with Ref.~\cite{Jiang2024}, consider the PFDM lapse
\begin{equation}
 f_{\rm PFDM}(r)=1-\frac{2M}{r}-\frac{b}{r}\ln\!\left(\frac{r}{b}\right)-\frac{\Lambda r^2}{3},
 \qquad b>0.
\end{equation}
At a cold double root $r_e$, write $b_e\equiv b_{\max}(\Lambda)$.  The conditions $f=f'=0$ give
\begin{equation}
 b_e=r_e(1-\Lambda r_e^2),\quad
 \frac{2M}{r_e}=1-\frac{\Lambda r_e^2}{3}
 +(1-\Lambda r_e^2)\ln(1-\Lambda r_e^2).
 \label{eq:PFDMcold}
\end{equation}
For $b=b_e(1-\epsilon_b)$, the double-root expansion yields
\begin{equation}
\begin{aligned}
 \kappam &=K_{\rm PFDM}\sqrt{\epsilon_b}+O(\epsilon_b),\\
 K_{\rm PFDM}&=\frac{1}{r_e}
 \sqrt{\frac{(1-\Lambda r_e^2)[1+\ln(1-\Lambda r_e^2)](1-3\Lambda r_e^2)}{2}}.
\end{aligned}
\label{eq:KPFDM}
\end{equation}
Since $b_e=b_{\max}$, define $\widehat b=b/b_{\max}$ and $\Delta\widehat b=1-\widehat b_{\rm crit}$.  The PFDM cold expansion then gives a direct analytic estimate of the PS-controlled violation width.  Writing $\bar r_{\rm ph}\equiv r_{e,{\rm ph}}/r_e>1$, the extremal photon-orbit condition reduces to
\begin{equation}
 2(\bar r_{\rm ph}-1)
 =3(1-\Lambda r_e^2)\ln \bar r_{\rm ph}.
 \label{eq:PFDMphoton}
\end{equation}
Using the standard eikonal relation between the PS damping rate and the photon-orbit Lyapunov exponent \cite{CardosoGeodesic2009}, one obtains
\begin{align}
 \lambda_{e,{\rm ph}}^2
 &=\frac{\bar r_{\rm ph}-1}{6r_e^2\bar r_{\rm ph}^4}
 \left[1-\Lambda r_e^2(\bar r_{\rm ph}^2+\bar r_{\rm ph}+1)\right]\nonumber\\
 &\quad\times\left(2\bar r_{\rm ph}+3\Lambda r_e^2-3\right).
 \label{eq:PFDMLyapunov}
\end{align}
Together with Eq.~\eqref{eq:KPFDM}, this gives the cold PS-width estimate
\begin{equation}
 \Delta\widehat b_{\rm PS}^{\rm cold}
 \simeq\left(\frac{\lambda_{e,{\rm ph}}}{K_{\rm PFDM}}\right)^2.
 \label{eq:PFDMPSwidth}
\end{equation}
The physical cold branch satisfies $0<\Lambda r_e^2<1/3$.  Since the second relation in Eq.~\eqref{eq:PFDMcold} determines $M/r_e$ as a function of $\Lambda r_e^2$, it also fixes $\Lambda M^2=(\Lambda r_e^2)(M/r_e)^2$, the parameter used in the numerical scan.  Along the physical cold branch, $\Lambda M^2$ increases whereas the estimate in Eq.~\eqref{eq:PFDMPSwidth} decreases monotonically.  Thus, once the PS family becomes the threshold controller, the PFDM cold PS-width estimate is already decreasing.

We first validate the PFDM implementation used for this comparison.  Replacing only the lapse function in our Chebyshev pseudospectral code by $f_{\rm PFDM}$ reproduces the QNM benchmarks of Ref.~\cite{Jiang2024}.  For example, at $\Lambda M^2=0.01$ we obtain $(\beta_{\ell=0},\beta_{\ell=1},\beta_{\rm PS}^{\rm WKB})=(0.745713,0.571284,0.552930)$ for $b/b_{\max}=0.99$ and $(0.869147,0.870278,0.842109)$ for $b/b_{\max}=0.995$, consistent with their Tables~1 and 2.

We then test directly the feature relevant to Sec.~\ref{sec:PFDMcomparison}.  A branch-resolved threshold scan brackets the dS--PS controller exchange by $0.00208<\Lambda M^2<0.00210$, and the global normalized width reaches its maximum in the same interval at this numerical resolution.  In the PS-controlled regime, both the finite-$\widehat b$ WKB threshold and the cold estimate decrease monotonically as $\Lambda M^2$ increases; representative values are listed in Table~\ref{tab:PFDMwidth}.  The leading cold approximation systematically overestimates the finite-$\widehat b$ widths, as expected away from the strict cold limit, but correctly reproduces the post-exchange narrowing.  Hence, in PFDM the maximum lies near the controller exchange because the PS-controlled branch is already narrowing when it becomes dominant, whereas in CDF the PS-controlled width continues to grow after the first exchange before reaching a separate maximum.

\begin{table}[!htbp]
\caption{PFDM PS-controlled normalized SCC-violation width from the finite-$\widehat b$ WKB threshold and from the cold estimate in Eq.~\eqref{eq:PFDMPSwidth}.}
\label{tab:PFDMwidth}
\begin{ruledtabular}
\begin{tabular}{c c c}
$\Lambda M^2$ & $\Delta\widehat b_{\rm PS}$ (WKB) & cold estimate\\
\hline
0.0021 & 0.01313 & 0.02273\\
0.0050 & 0.01267 & 0.02193\\
0.0100 & 0.01172 & 0.02029\\
0.0200 & 0.00886 & 0.01534\\
0.0300 & 0.00230 & 0.00412\\
\end{tabular}
\end{ruledtabular}
\end{table}

\begin{acknowledgments}
We thank J. Jiang and S. Jiang for helpful discussions. This work was supported by the National Natural Science Foundation of China under Grant No.~12305070, and the Basic Research Program of Shanxi Province under Grant Nos.~202303021222018 and 202303021221033.
\end{acknowledgments}

\bibliographystyle{apsrev4-2-titles}
\bibliography{references}

\end{document}